\documentclass[journal]{IEEEtran}

\usepackage{cite}
\usepackage{amsmath,amssymb,amsfonts}
\usepackage{algorithm}  
\usepackage{algpseudocode}  
\usepackage{algpseudocode}
\usepackage{graphicx}
\usepackage{epstopdf}
\usepackage{textcomp}
\usepackage{xcolor}
\usepackage{bm}
\usepackage{setspace}

\ifCLASSINFOpdf
 
\else
 
\fi

\begin{document}

\title{Beamforming Design and Power Allocation for Transmissive RMS-based Transmitter Architectures}

\author{Zhendong~Li,~Wen~Chen,~\IEEEmembership{Senior~Member,~IEEE},~and~Huanqing~Cao
	\thanks{Z. Li, W. Chen and Q. Cao are with the Department of Electronic Engineering, Shanghai Jiao Tong University, Shanghai 200240, China (e-mail: lizhendong@sjtu.edu.cn; wenchen@sjtu.edu.cn; caohuanqing@sjtu.edu.cn).}
	\thanks{(\emph{Corresponding author: Wen Chen.})}}

\maketitle

\begin{abstract}
This letter investigates a downlink multiple input single output (MISO) system based on transmissive reconfigurable metasurface (RMS) transmitter. Specifically, a transmitter design based on a transmissive RMS equipped with a feed antenna is first proposed. Then, in order to maximize the achievable sum-rate of the system, the beamforming design and power allocation are jointly optimized. Since the optimization variables are coupled, this formulated optimization problem is non-convex, so it is difficult to solve it directly. To solve this problem, we propose an alternating optimization (AO) technique based on difference-of-convex (DC) programming and successive convex approximation (SCA). Simulation results verify that the proposed algorithm can achieve convergence and improve the achievable sum-rate of the system.
\end{abstract}


\IEEEpeerreviewmaketitle

\section{Introduction}
The power consumption of a single base station (BS) for 5G is 3 to 4 times that of 4G. Meanwhile, since 5G uses a small-cell network, its power consumption per unit area will increase by 3 times. In addition, 5G BSs equipped with large-scale antenna arrays require a large number of radio frequency (RF) chains and complex signal processing technologies, power consumption and deployment cost of networks are greatly increased. Therefore, it is essential for B5G and 6G to seek a novel network paradigm that can reduce power consumption and cost.

Reconfigurable metasurface (RMS), also known as reconfigurable intelligent surface (RIS), has been proposed as a revolutionary technology, which can be used to improve network coverage, spectrum- and energy- efficiency. RMS is an array composed of a large number of low-cost passive elements. By adjusting these elements, the amplitude and phase shift of the incident electromagnetic (EM) wave can be changed, and then the wireless channel can be reconstructed. Compared with the multi-antenna technology, the required hardware cost and power consumption are much lower \cite{li2021transmissive}. These have greatly promoted the application of RMS in B5G and 6G networks.

With the above advantages, RMS has attracted widespread attention in the field of wireless communications. Nowadays, RMS is mainly used to assist communication networks. By joint active and passive beamforming design, the performance of the network can be improved by relying on a reflective or transmissive RMS \cite{9531372,9167258,9365004,zhang2021intelligent,liu2021star}. First, some research has been carried out on the reflective RMS-assisted communication networks. Zuo \emph{et al.} maximized the throughput for the intelligent reflecting surface (IRS)-assisted non-orthogonal multiple access (NOMA) networks \cite{9167258}. Mu \emph{et al.} studied the fundamental capacity limitation of IRS-assisted multi-user wireless communication system \cite{9365004}.

In addition, some researches on transmissive RMS are also in progress. Zhang \emph{et al.} proposed the concept of intelligent omni-surface (IOS), which can serve mobile users on both sides of the surface and realize multi-dimensional communication through its reflection and transmission characteristics. Meanwhile, the working principle of IOS is also introduced, and a novel hybrid beamforming scheme is proposed for wireless communication based on IOS \cite{zhang2021intelligent}. Liu \emph{et al.} proposed a novel simultaneously transmitting and reflecting (STAR) system which relies on RMS \cite{liu2021star}. In addition to assisting communication networks, RMS's reconfigurability helps it expand the number of passive elements without increasing the number of expensive active antennas. Moreover, the reflective RMS equipped with a feed antenna can also be used as a transmitter \cite{9133266}. However, the reflective RMS is less efficient for two main reasons. First, the reflective RMS has a more serious feed blockage compared to the transmitted RMS. Second, the reflective RMS has the occlusion of the feed source.  \cite{bai2020high} has demonstrated that the aperture efficiency and working bandwidth of transmissive RMS is higher than those of reflective RMS.

Inspired by the aforementioned works, a transmitter design based on a more efficient transmissive RMS is proposed in this paper. Such a transmissive RMS-based transmitter can greatly solve the problems of high power consumption and high cost for 5G networks, thus it is a very promising architecture. As far as we know, there is currently no existing research on beamforming design and power allocation under this architecture. In this paper, we first propose a transmitter design based on a transmissive RMS equipped with a feed antenna. Then, the system achievable sum-rate maximization problem of joint beamforming design and power allocation is formulated. Due to the high coupling of optimization variables, it is non-trivial to solve this optimization problem. We propose a joint alternating optimization (AO) optimization algorithm based on difference-of-convex (DC) programming and successive convex approximation (SCA) to solve it.

\section{Transmissive RMS-based Transmitter Design}
In this section, we introduce a transmitter architecture based on a transmissive RMS equipped with a feed antenna. Specifically, we propose to perform transmissive RMS-based quadrature phase shift keying (QPSK) modulation and beamforming design by adjusting the amplitude and phase shift of RMS elements in real time. QPSK modulation can be achieved by deploying two diodes in each element \cite{9133266}. It is worth noting that this paper uses QPSK as an example to illustrate the transmitter design based on transmissive RMS. Other higher-order modulation methods can also be implemented, such as quadrature amplitude modulation (16QAM) \cite{9133266}. Next, we further elaborate transmissive RMS-based transmitter and QPSK signal moduation mapping. Herein, the feed antenna transmits single-frequency EM wave. The information source bits from signal source can be mapped to corresponding phase shift, denoted by $e^{j\theta}$, as one part transmissive coefficient of RMS\footnote{This paper mainly proposes a new transmitter architecture, and then performs beamforming design and power allocation under this architecture. Therefore, we only consider the ideal case where the amplitude and phase of RMS are assumed to be independently adjustable. In fact, there is a certain relationship between the amplitude and phase of the RMS \cite{9115725}, and we will leave this more practical model as future work.}. The other part transmissive coefficient denoted by $f=\beta e^{j\theta'}$ is obtained via beamforming design algorithm described later. Then the ultimate transmissive coefficient can be expressed as $\beta e^{j\theta'}e^{j\theta}$. Then RMS controller can adjust the transmissive coefficient of each element by setting corresponding control signal to realize the signal moduation and transmission. The key steps can be summarized as: (a) Get the information bit streaming (0100111...) from the signal source. (b) Map the bit stream into the corresponding QPSK phase shift. (c) Obtain the beamforming design according to proposed beamforming algorithm. (d) Acquire the transmissive coefficient of RMS according to step (b) and step (c) and determine the control signal of RMS. (e) Control the transmissive coefficient of RMS according to the control signal obtained in step (d) and then the transmissive EM wave modulated with the information is transmitted once the incident EM wave arrives the RMS.
\section{System Model and Problem Formulation}
In this section, we mainly propose the beamforming algorithm in step (c) above-mentioned. We first introduce the multiple input single output (MISO) system based on transmissive RMS transmitter. As shown in Fig. 1, the network is composed of a feed antenna that constantly radiates EM waves (i.e., single-tone carrier signals), a transmissive RMS equipped with $M$ elements, and $K$ single-antenna users. The transmitter consists of a feed antenna and a transmissive RMS, and the total transmit power is ${P_t}$. According to the above-mentioned transmitter design, the transmitter maps the information from signal source to be sent to the transmission control signal of the RMS to realize the modulation of the information signal and the beamforming of the transmissive EM wave. Note that the transmitter sends a single carrier QPSK modulated signal at the transmitter. Let ${\bf{f}} = {\left[ {{f_1},...,{f_m}} \right]^T} \in\mathbb{C} {^{M \times 1}}$ denote the signal transmission coefficient of RMS in the downlink (DL), ${f_m} = {\beta _m}{e^{j{\theta _m}}},\forall m$, where ${\beta _m} \in \left[ {0,1} \right]$ denotes the amplitude of transmissive EM wave, and ${\theta _m} \in \left[ {0,2\pi } \right)$ denotes the phase shift of transmissive EM wave. There are the following constraints on the transmission coefficient ${f_m}$,
\begin{equation}
    \left| {{f_m}} \right| \le 1,\forall m.
\end{equation}
In order to facilitate the analysis, we consider the case where the signal is all transmitted, i.e., no signal is reflected.
\begin{figure}
	\centerline{\includegraphics[width=8.5cm]{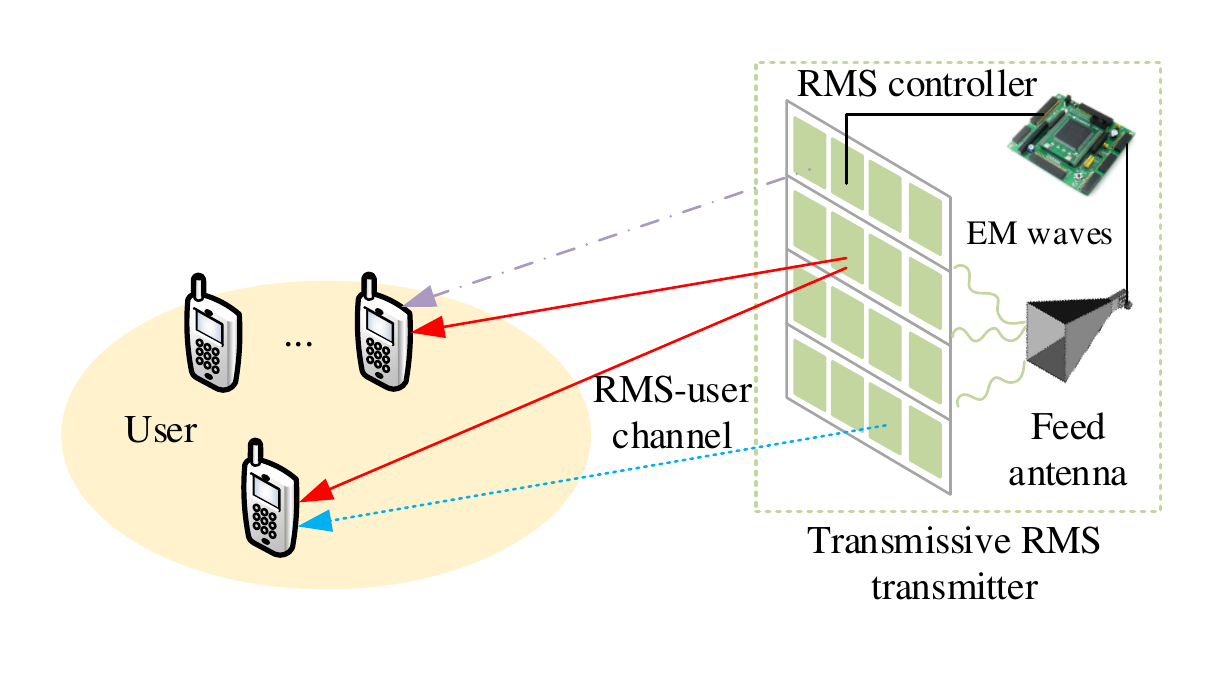}}
	\caption{MISO system based on transmissive RMS transmitter.}
\end{figure}
For DL, the channel gain from the transmitter to the $k$-th user can be expressed as ${\bf{h}}_{t,k}^H \in \mathbb{C}{}{^{1 \times M}}$, and we assume that all channels are quasi-static and flat fading. Moreover, in order to characterize the best theoretical performance under our proposed architecture, we assume that the channel state information (CSI) can be perfectly obtained at RMS controller. It is worth noting that we consider time division duplex (TDD) protocol for uplink (UL) and DL transmission. Specifically, for the UL, users send pilots to the receiving end, and different observation matrices can be obtained by randomly switching different RMS transmissive coefficients to estimate the channel, thereby obtaining the CSI of the UL. The CSI of the DL can also be obtained according to the reciprocity of the channels. There are also some investigations on RMS channel estimation \cite{8879620}. In this paper, we consider a more practical array response, i.e., the RMS array is distributed in a uniform planar array (UPA). Let $M = {M_x} \times {M_z}$, ${M_x}$ and ${M_z}$ respectively represent the number of horizontal and vertical transmissive RMS elements. We use a geometric channel model with $L$ propagation paths to model the channel between the transmitter and the user, which can be expressed as
\begin{equation}
    {{\bf{h}}_{t,k}} = \sqrt {\frac{M}{L}} \sum\limits_{l = 1}^L {{\xi _{k,l}}{\bf{a}}\left( {\theta _{k,l}^{{\rm{AoD}}},\phi _{k,l}^{{\rm{AoD}}}} \right)} ,\forall k,
\end{equation}
where $L$ is the number of propagation paths, ${\xi _{k,l}}$ is the complex channel gain, $\theta _{k,l}^{{\rm{AoD}}}$ and $\phi _{k,l}^{{\rm{AoD}}}$ are the azimuth angle and elevation angle of the angle-of-departure (AoD) at the transmissive RMS, respectively. ${\bf{a}}\left( {\theta _{k,l}^{{\rm{AoD}}},\phi _{k,l}^{{\rm{AoD}}}} \right)$ is the array response at the transmissive RMS, which is determined by Eq. (3) and is shown at the top of the next page. In Eq. (2), $l = 1$ represents the LoS path, and $l > 1$ represents the NLoS path.
\newcounter{mytempeqncnt}
\begin{figure*}[!t]
	\normalsize
	\setcounter{equation}{2}
	\begin{small}
	\begin{equation}
		{\bf{a}}\!\left( {\theta _{k,l}^{{\rm{AoD}}}\!,\!\phi _{k,l}^{{\rm{AoD}}}} \right) \!\!=\!\! \frac{1}{{\sqrt M }}{\!\left[ {1,\!...\!,{e^{ \!- \!j\frac{{2\pi }}{\lambda }d\left( {{m_x}\sin \left( {\theta _{k,l}^{{\rm{AoD}}}} \right)\sin \left( {\phi _{k,l}^{{\rm{AoD}}}} \right) \!+\! {m_z}\cos \left( {\phi _{k,l}^{{\rm{AoD}}}} \right)} \right)}}\!,...,\!{e^{ \!-\! j\frac{{2\pi }}{\lambda }d\left( {\left( {{M_x}\! -\! 1} \right)\sin \left( {\theta _{k,l}^{{\rm{AoD}}}} \right)\sin \left( {\phi _{k,l}^{{\rm{AoD}}}} \right) \!+ \!\left( {{M_z} \!-\! 1} \right)\cos \left( {\phi _{k,l}^{{\rm{AoD}}}} \right)} \right)}}} \right]^T}\!,\!\forall k,\!l,
	\end{equation}
	\end{small}
\hrulefill
\vspace*{4pt}
\end{figure*}

Let ${s_k}$ denote the signal sent by the transmitter to the $k$-th user, which is assumed to be independent and identically distributed (i.i.d) CSCG random variable with zero mean and unit variance, i.e., ${s_k} \sim {\cal C}{\cal N}\left( {0,1} \right)$. Let ${p_k}$ denote the transmit power allocation for the $k$-th user, which satisfies the following constraints
\begin{equation}
    {p_k} > 0,\forall k,~~\sum\limits_{k = 1}^K {{p_k} \le {P_t}} .
\end{equation}
%
In DL, the signal received by the $k$-th user can be expressed as
\begin{equation}
    {y_k} = {\bf{h}}_{t,k}^H{\bf{f}}\sqrt {{p_k}} {s_k} + {\bf{h}}_{t,k}^H{\bf{f}}\sum\limits_{i \ne k} {\sqrt {{p_i}} {s_i} + {n_k}} ,\forall k,
\end{equation}
where $n_k$ represents the additive white Gaussian noise (AWGN) introduced by the receiver, ${n_k} \sim {\cal C}{\cal N}\left( {0,\sigma _k^2} \right)$. Therefore, the signal to interference and noise ratio (SINR) for the $k$-th user can be expressed as
\begin{equation}
    {\rm{SIN}}{{\rm{R}}_k} = \frac{{{p_k}{{\left| {{\bf{h}}_{t,k}^H{\bf{f}}} \right|}^2}}}{{\sum\limits_{i \ne k} {{p_i}{{\left| {{\bf{h}}_{t,k}^H{\bf{f}}} \right|}^2} + \sigma _k^2} }},\forall k.
\end{equation}
Therefore, the achievable rate of the $k$-th user is given by
\begin{equation}
    {R_k}{\rm{ = lo}}{{\rm{g}}_2}\left( {1 + {\rm{SIN}}{{\rm{R}}_k}} \right),\forall k.
\end{equation}
Next, the achievable sum-rate of the system can be given by
\begin{equation}
    {R_{sum}}{\rm{ = }}\sum\limits_{k = 1}^K {{R_k}.} 
\end{equation}

Let ${\bf{p}} = \left[ {{p_1},...,{p_K}} \right]$ denote transmit power allocation. We maximize the achievable sum-rate of the system by jointly optimizing the transmission coefficient ${\bf{f}}$ of the transmissive RMS and the transmit power allocation ${\bf{p}}$. The optimization problem can be expressed as follows
\begin{subequations}
	\begin{align}
		\left( {{\textrm{P1}}} \right){\rm{~~~~~}}&\mathop {\max }\limits_{{\bf{f}},{\bf{p}}} {\rm{~}}{R_{sum}}, \\
		\rm{s.t.}\qquad &{\rm{SIN}}{{\rm{R}}_k} \ge {\gamma _{th}},\forall k,\\
		&\left| {{f_m}} \right| \le 1,\forall m,\\
		&{p_k} > 0,\forall k,\sum\limits_{k = 1}^K {{p_k} \le {P_t}} ,
	\end{align}
\end{subequations}
where constraint (9b) is the user's QoS requirement constraint. The constraint (9c) is the transmission coefficient constraint of the transmissive RMS, and the constraints (9d) is the transmit power constraints. Obviously, since the two optimization variables are coupled, the objective function is a non-concave function, and the constraint (9b) is a non-convex constraint. Therefore, the problem is a non-convex optimization problem, which is non-trivial to solve directly.
\section{Proposed Algorithm}
To solve the non-convex optimization problem (P1), we apply alternating optimization (AO) algorithm. Specifically, we first decouple it into two sub-problems, i.e., RMS transmission coefficient optimization and transmit power allocation. Then we alternately optimize the two sub-problems until convergence is achieved. These two sub-problems can be expressed as problem (P2) and problem (P3) respectively as follows
\begin{subequations}
	\begin{align}
		\left( {{\textrm{P2}}} \right){\rm{~~~~~}}&\mathop {\max }\limits_{{\bf{f}}} {\rm{~}}{R_{sum}}, \\
		\rm{s.t.}\qquad &\textrm {(9b), (9c)} ,
	\end{align}
\end{subequations}
and 
\begin{subequations}
	\begin{align}
		\left( {{\textrm{P3}}} \right){\rm{~~~~~}}&\mathop {\max }\limits_{{\bf{p}}} {\rm{~}}{R_{sum}}, \\
		\rm{s.t.}\qquad &\textrm {(9b), (9d)}.
	\end{align}
\end{subequations}
The achievable rate of the $k$-th user can also be expressed as
\begin{small}
\begin{equation}
    {R_k}{\rm{ = }}{\log _2}\left( {\sum\limits_{i = 1}^K {{p_i}{{\left| {{\bf{h}}_{t,k}^H{\bf{f}}} \right|}^2} \!+\! \sigma _k^2} } \right) \!-\! {\log _2}\left( {\sum\limits_{i \ne k} {{p_i}{{\left| {{\bf{h}}_{t,k}^H{\bf{f}}} \right|}^2}\! +\! \sigma _k^2} } \right),\forall k,
\end{equation}
\end{small}Let ${{\bf{H}}_{t,k}} = {{\bf{h}}_{t,k}}{\bf{h}}_{t,k}^H$ and ${\bf{F}} = {\bf{f}}{{\bf{f}}^H}$, where ${{\bf{H}}_{t,k}} \succeq 0$, ${\rm{rank}}\left( {{{\bf{H}}_{t,k}}} \right) = 1$, ${\bf{F}} \succeq 0$ and ${\rm{rank}}\left( {\bf{F}} \right) = 1$, then the achievable rate of the $k$-th user can be further expressed as
\begin{small}
\begin{equation}
    {R_k}{\rm{ \!=  \!}}{\log _2} \!\left( {\sum\limits_{i = 1}^K {{p_i}{\rm{tr}}\left( {{\bf{F}}{{\bf{H}}_{t,k}}} \right)  \! \!+ \! \! \sigma _k^2} } \right) \!  - \!  {\log _2} \!\left( \! {\sum\limits_{i \ne k} {{p_i}{\rm{tr}}\left( \! {{\bf{F}}{{\bf{H}}_{t,k}}} \right)  \! \!+ \! \! \sigma _k^2} } \right),\forall k.
\end{equation}
\end{small}
\subsection{RMS transmission coefficients optimization}
In this subsection, given the transmit power allocation, we solve for the problem (P2), i.e., RMS transmission coefficient optimization. More specifically, Eq. (13) can be expressed as a function of ${\bf{F}}$ as follows
\begin{equation}
    {R_k}{\rm{ = }}{g_1}\left( {\bf{F}} \right) - {g_2}\left( {\bf{F}} \right),\forall k.
\end{equation}
Since ${R_k}$ is the difference between two concave functions of the matrix $\textbf{F}$, it is not a concave function, so the problem P2 is not a convex optimization problem. Then, we use SCA\footnote{Note that any convex function is subject to the global lower bound of its first-order Taylor expansion at any point, and any concave function is subject to the global upper bound of its first-order Taylor expansion at any point.} to linearly approximate the second term of the objective function (i.e., ${g_2}\left( {\bf{F}} \right)$) as follows
\begin{equation}
	\begin{aligned}
		 {g_2}\left( {\bf{F}} \right) &\le {g_2}\left( {{{\bf{F}}^{\left( r \right)}}} \right) + {\rm{tr}}\left( {{{\left( {{\nabla _{\bf{F}}}{g_2}\left( {{{\bf{F}}^{\left( r \right)}}} \right)} \right)}^H}\left( {{\bf{F}} - {{\bf{F}}^{\left( r \right)}}} \right)} \right)\\
		& \buildrel \Delta \over = {g_2}{\left( {\bf{F}} \right)^{ub}},\forall k,
	\end{aligned}
\end{equation}
where ${{\bf{F}}^{\left( r \right)}}$ denotes the value of $\bf{F}$ in the $r$-th iteration, and
\begin{equation}
    {\nabla _{\bf{F}}}{g_2}\left( {{{\bf{F}}^{\left( r \right)}}} \right) = \frac{{\sum\limits_{i \ne k} {{p_i}{\bf{H}}_{t,k}^H} }}{{\left( {\sum\limits_{i \ne k} {{p_i}{\rm{tr}}\left( {{{\bf{F}}^{\left( r \right)}}{{\bf{H}}_{t,k}}} \right) + \sigma _k^2} } \right)\ln 2}},\forall k.
\end{equation}
Therefore, the problem (P2) can be further approximately transformed into the problem (P2.1), which can be given by
\begin{subequations}
	\begin{align}
		\left( {{\textrm{P2.1}}} \right){\rm{~~~~~}}&\mathop {\max }\limits_{\bf{F}} {\rm{ }}\sum\limits_{k = 1}^K {\left( {{g_1}\left( {\bf{F}} \right) - {g_2}{{\left( {\bf{F}} \right)}^{ub}}} \right)} , \\
		\rm{s.t.}\qquad &{p_k}{\rm{tr}}\left( {{\bf{F}}{{\bf{H}}_{t,k}}} \right) \ge {\gamma _{th}}\left( {\sum\limits_{i \ne k} {{p_i}{\rm{tr}}\left( {{\bf{F}}{{\bf{H}}_{t,k}}} \right) + \sigma _k^2} } \right),\forall k,\\
		&{{\bf{F}}_{m,m}} \le 1,\forall m,\\
		&{\bf{F}} \succeq 0,\\
		&{\rm{rank}}\left( {\bf{F}} \right) = 1.
	\end{align}
\end{subequations}
Due to the non-convex rank-one constraint, the problem (P2.1) is still a non-convex optimization problem. Next, we use the DC programming to deal with this rank-one constraint.

$\textbf{Proposition 1:}$ For the positive semi-definite matrix ${\bf{F}} \in {\mathbb{C}^{M \times M}}$, ${\rm{tr}}\left( {\bf{F}} \right) > 0$, the rank-one constraint can be expressed as the difference between two convex functions, i.e.,
\begin{equation}
	{\rm{rank}}\left( {\bf{F}} \right) = 1 \Leftrightarrow {\rm{tr}}\left( {\bf{F}} \right) - {\left\| {\bf{F}} \right\|_2}=0,
\end{equation}
where ${\rm{tr}}\left( {\bf{F}} \right) = \sum\limits_{n = 1}^M {{\sigma _n}\left( {\bf{F}} \right)} $, ${\left\| {\bf{F}} \right\|_2} = {\sigma _1}\left( {\bf{F}} \right)$ is spectral norm, and ${\sigma _n}\left( {\bf{F}} \right)$ represents the $n$-th largest singular value of matrix ${\bf{F}}$ \cite{greub2012linear}.

According to $\textbf{Proposition 1}$, we transform the non-convex rank-one constraint on matrix ${\bf{F}}$, and then add it as a penalty term to the objective function of problem (P2.1). Therefore, problem (P2.1) can be transformed into the problem (P2.2) as follows
\begin{subequations}
	\begin{align}
		\left( {{\textrm{P2.2}}} \right){\rm{~~~~~}}&\mathop {\max }\limits_{\bf{F}} {\rm{ }}\sum\limits_{k = 1}^K {\left( {{g_1}\left( {\bf{F}} \right) - {g_2}{{\left( {\bf{F}} \right)}^{ub}}} \right)}  - \varsigma \left( {{\rm{tr}}\left( {\bf{F}} \right) - {{\left\| {\bf{F}} \right\|}_2}} \right), \\
		\rm{s.t.}\qquad &\textrm {(17b), (17c), (17d)}, 
	\end{align}
\end{subequations}
where $\varsigma  \gg 0$ represents the penalty factor related to the rank-one constraint. Since ${\left\| {\bf{F}} \right\|_2}$ is a convex function with respect to $\bf{F}$, the problem (P2.2) is still a non-convex optimization problem. Herein, we use SCA to perform a first-order Taylor expansion of ${\left\| {\bf{F}} \right\|_2}$ to obtain its lower bound as follows
\begin{equation}
	\begin{aligned}
		 {\left\| {\bf{F}} \right\|_2}&\! \ge\! {\left\| {{{\bf{F}}^{\left( r \right)}}} \right\|_2} \!+\! {\rm{tr}}\!\left( \!{{{\bf{u}}_{\max }}\!\left(\! {{{\bf{F}}^{\left( r \right)}}} \right)\!{{\bf{u}}_{\max }}{{\left( {{{\bf{F}}^{\left( r \right)}}} \right)}^H}\left( \!{{\bf{F}} \!-\! {{\bf{F}}^{\left( r \right)}}} \right)}\! \right)\\
		& \buildrel \Delta \over = {\left( {{{\left\| {\bf{F}} \right\|}_2}} \right)^{lb}},
	\end{aligned}
\end{equation}
where ${{\bf{u}}_{\max }}\left( {{{\bf{F}}^{\left( r \right)}}} \right)$ represents the eigenvector corresponding to the largest singular value of matrix $\bf{F}$ in the $r$-th iteration. Therefore, the problem (P2.2) can be approximately transformed into a problem (P2.3), which can be expressed as
\begin{subequations}
	\begin{align}
		\left( {{\textrm{P2.3}}} \right){\rm{~~~}}&\mathop {\max }\limits_{\bf{F}} {\rm{ }}\sum\limits_{k = 1}^K {\left( {{g_1}\left( {\bf{F}} \right) \!- \! {g_2}{{\left( {\bf{F}} \right)}^{ub}}} \right)}   \!- \! \varsigma \left( {{\rm{tr}}\left( {\bf{F}} \right) \! - \! {{\left( {{{\left\| {\bf{F}} \right\|}_2}} \right)}^{lb}}} \right), \\
		\rm{s.t.}\qquad &\textrm {(17b), (17c), (17d)}.
	\end{align}
\end{subequations}
It can be seen that this problem is a standard semi-definite programming (SDP) problem, which can be solved by using CVX toolbox to obtain the transmission coefficient of the transmissive RMS \cite{cvx}.
\subsection{Transmit power allocation}
In this subsection, based on the obtained RMS transmission coefficient, we obtain the transmit power allocation. Furthermore, Eq. (13) can be expressed as a function of ${p_i}$ as follows,
\begin{equation}
    {R_k}{\rm{ = }}{h_1}\left( {{p_i}} \right) - {h_2}\left( {{p_i}} \right),\forall k.
\end{equation}
Since ${R_k}$ is the difference between two concave functions, it is non-concave. We use SCA to perform a first-order Taylor expansion on ${h_2}\left( {{p_i}} \right)$ as follows
\begin{equation}
	\begin{aligned}
		 &{h_2}\left( {{p_i}} \right) \le\\
		 &{h_2}\left( {{p_i}^{\left( r \right)}} \right) + \frac{{{\rm{tr}}\left( {{\bf{F}}{{\bf{H}}_{t,k}}} \right)}}{{\left( {\sum\limits_{i \ne k} {{p_i}^{\left( r \right)}{\rm{tr}}\left( {{\bf{F}}{{\bf{H}}_{t,k}}} \right) + \sigma _k^2} } \right)\ln 2}}\left( {{p_i} - {p_i}^{\left( r \right)}} \right)\\
		& \buildrel \Delta \over = {h_2}{\left( {{p_i}} \right)^{ub}},\forall k,
	\end{aligned}
\end{equation}
where ${p_i}^{\left( r \right)}$ denotes the value of ${p_i}$ of the $r$-th iteration. Therefore, the problem (P3) can be approximately transformed into the problem (P3.1), which can be expressed as
\begin{subequations}
	\begin{align}
		\left( {{\textrm{P3.1}}} \right){\rm{~~~~~}}&\mathop {\rm{ }}\mathop {\max }\limits_{\bf{p}} {\rm{ }}\sum\limits_{k = 1}^K {\left( {{h_1}\left( {{p_i}} \right) - {h_2}{{\left( {{p_i}} \right)}^{ub}}} \right)} , \\
		\rm{s.t.}\qquad &{p_k}{\rm{tr}}\left( {{\bf{F}}{{\bf{H}}_{t,k}}} \right) \ge {\gamma _{th}}\left( {\sum\limits_{i \ne k} {{p_i}{\rm{tr}}\left( {{\bf{F}}{{\bf{H}}_{t,k}}} \right) + \sigma _k^2} } \right),\forall k,\\
		&{p_k} > 0,\forall k,\sum\limits_{k = 1}^K {{p_k} \le {P_t}} .
	\end{align}
\end{subequations}
It can be seen that this problem is a standard convex optimization problem, which can be solved by using the CVX toolbox to obtain the transmit power allocation \cite{cvx}. The proposed joint beamforming design and power allocation optimization algorithm can be summarized as $\textbf{Algorithm 1}$\footnote{In each iteration, the problem (P2.3) solves a relaxed SDP problem by interior point method, so the computational complexity can be represented by ${\cal O}\left( {{M^{3.5}}} \right)$. The problem (P3.1) is solved with the complexity of ${\cal O}\left( K \right)$. We assume that the number of iterations required for the algorithm to reach convergence is $r$, the computational complexity of the proposed algorithm can be expressed as ${\cal O}\left( {r\left( {K + {M^{3.5}} } \right)} \right)$. In each iteration of $\textbf{Algorithm 1}$, the objective function of the problem (P1) is monotonically non-decreasing. Due to the limitation of transmit power, the achievable sum-rate of the system has an upper bound. Therefore, the convergence of the proposed algorithm can be guaranteed.} .

\begin{figure*}[htbp]
	\centering
	\begin{minipage}[c]{0.3\textwidth} 
		\centering
		\includegraphics[width=0.91\textwidth]{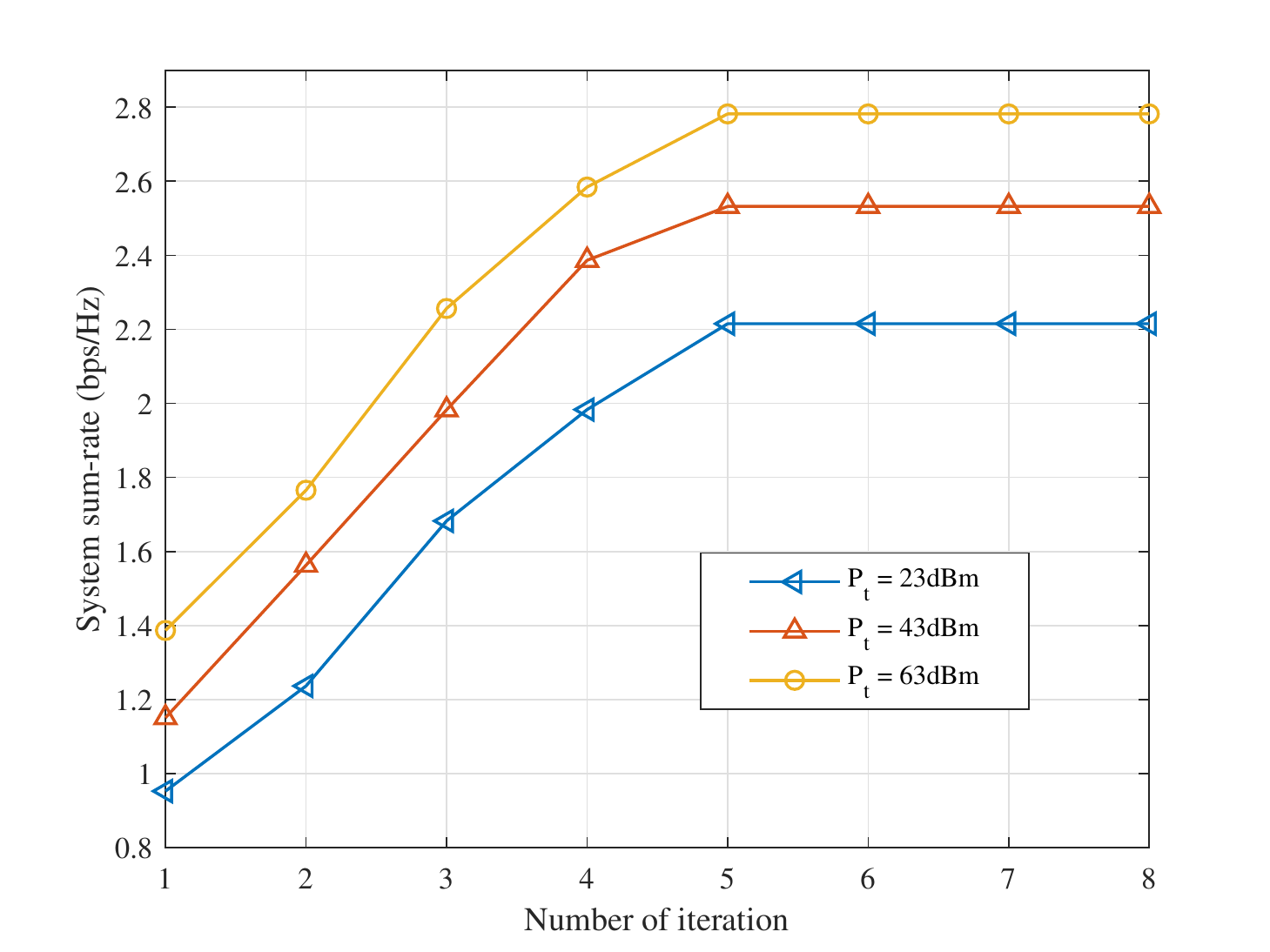} 
		\caption{Convergence behaviour of the proposed optimization algorithm.}
	\end{minipage}%
	\begin{minipage}[c]{0.3\textwidth}
		\centering
		\includegraphics[width=0.91\textwidth]{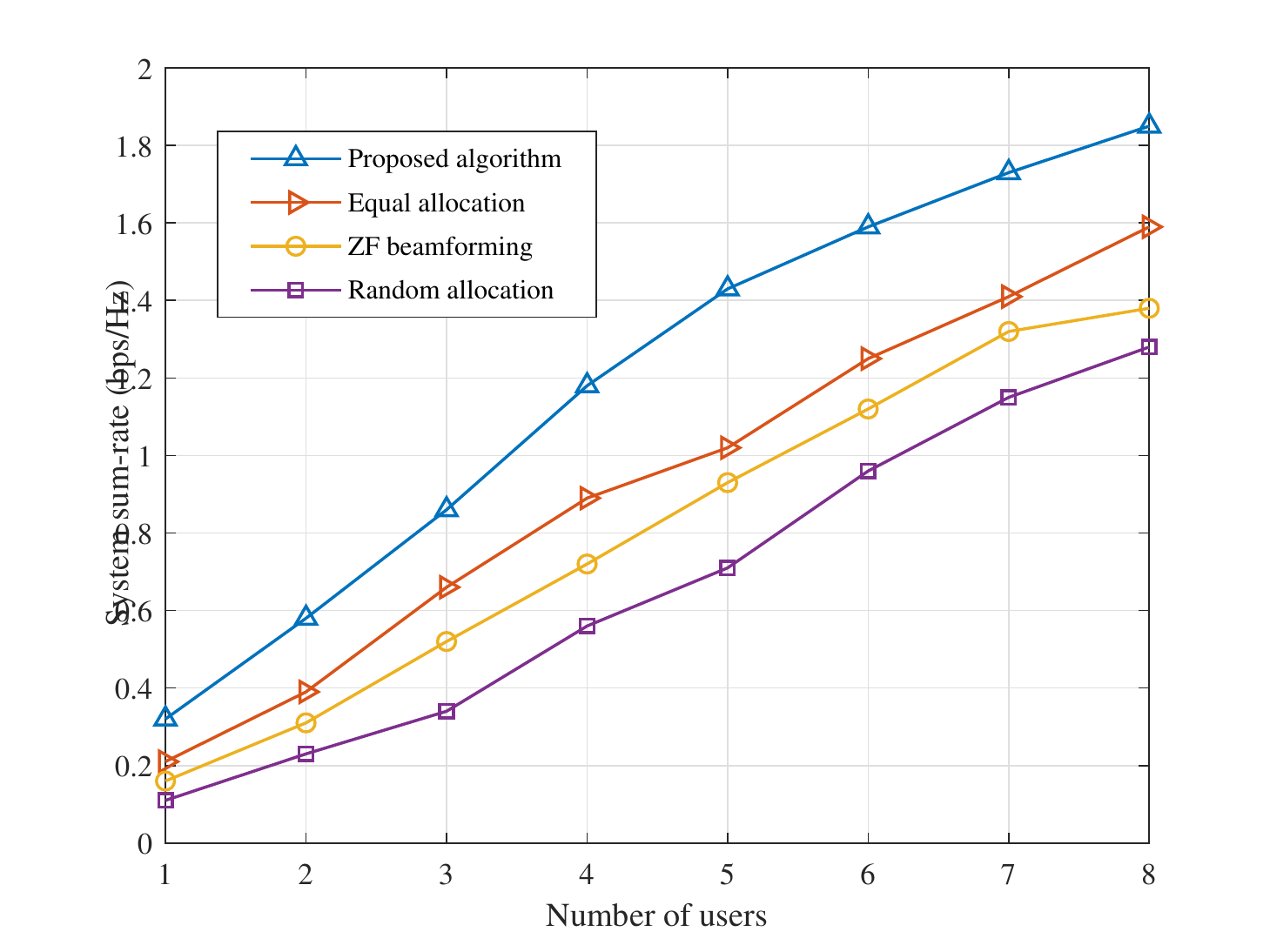}
		\caption{System sum-rate versus number of users under different algorithm.}
	\end{minipage}
	\begin{minipage}[c]{0.3\textwidth}
		\centering
		\includegraphics[width=0.91\textwidth]{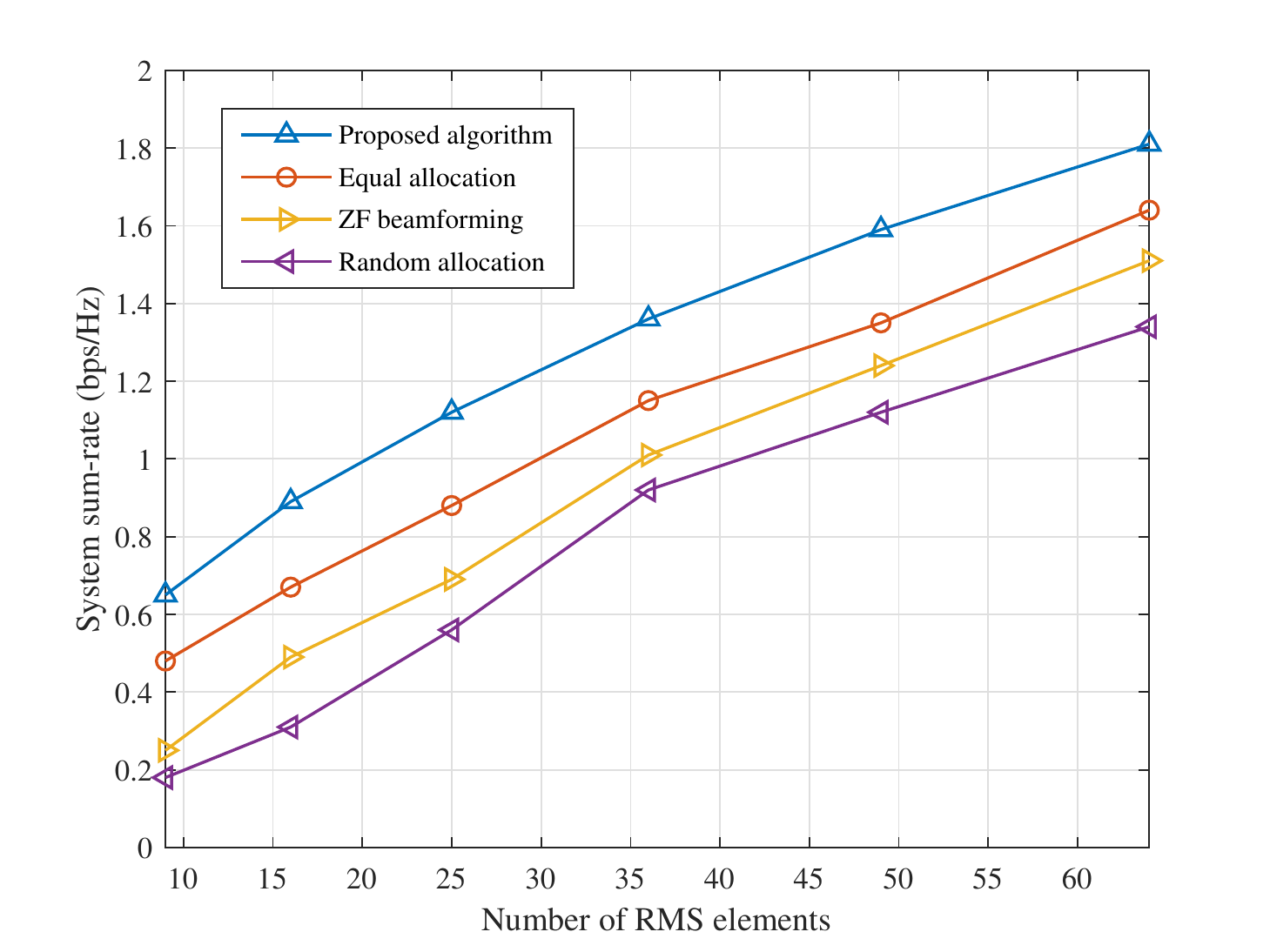}
		\caption{System sum-rate versus number of RMS elements under different algorithm.}
	\end{minipage}	
\end{figure*}

\begin{algorithm}
	\caption{Joint Beamforming Design and Power Allocation Algorithm} 
	\begin{algorithmic}[1]
		\Repeat
		\State Solve the problem (P2.3) for given ${{\bf{p}}^{\left( r \right)}}$, and obtain RMS transmission coefficients ${{\bf{f}}^{\left( r + 1 \right)}}$.
		\State Solve the problem (P3.1) for given ${{\bf{f}}^{\left( r \right)}}$, and obtain transmit power allocation ${{\bf{p}}^{\left( r + 1 \right)}}$.
		\State Update $r=r+1$.
		\Until  Until convergence is achieved.
		\State \Return Beamforming design and power allocation.
	\end{algorithmic}
\end{algorithm}
%

\section{Numercial Results}
In this section, we provide simulation results to demonstrate the effectiveness of the proposed algorithm. In this paper, we consider a three-dimensional (3D) coordinate system, where the RMS is located at (0m, 0m, 15m), and $K=4$ users are randomly and uniformly distributed in a circle whose origin is (0m, 0m, 0m) and radius is 50m \cite{9167258}. We consider that the RMS is equipped with 25 elements. Set the number of propagation paths $L = 3$. We assume that the parameters of all users are the same. Herein, we set $\sigma ^2=-70$dBm and $P_{t}=43$dBm. The convergence threshold of the proposed algorithm is set as $10^{-3}$. 

We first evaluate the convergence performance of the proposed algorithm. Fig. 2 depicts the system sum-rate varies with the number of iterations under different $P_{t}$. It can be seen that as the number of iterations increases, the system sum-rate gradually increases. The algorithm can achieve convergence in the 5-th iteration, which verifies that the proposed algorithm converges fast. It can also be seen that by increasing total transmit power, the system sum-rate is also increasing, which is consistent with the multi-antenna system.

Next, we compare the performance of the proposed algorithm with other baseline algorithms in Fig. 3. (1) Equal allocation (EA): The beamforming design is the same as our proposed algorithm, and the power allocation is equal allocation. (2) Zero-forcing (ZF) beamforming: The beamforming design applies ZF algorithm, and the power allocation is equal allocation. (3) Random allocation (RA): The beamforming design and power allocation adopt random allocation. As the number of users increases, the system sum-rate continues to increase under different algorithms. Meanwhile, when the number of users is the same, the performance of our proposed algorithm is better than the other algorithms. This is mainly because EA only considers the beamforming design, the power is distributed equally, the ZF beamforming design is not optimal, and the RA applies random allocation.

Finally, Fig. 4 describes the change of system sum-rate with the number of RMS elements. It can be seen that as the number of RMS elements increases, the system sum-rate increases. This is because more RMS elements can bring more spatial diversity gains. Meanwhile, more RMS elements can get more signal scattering paths, so that the receiving end can get a stronger signal. In addition, when the number of RMS elements is the same, our proposed algorithm has a better gain than other baseline algorithms.

\section{Conclusion}
In this paper, we have proposed transmissive RMS-based MISO transmitter architecture. Based on this architecture, the joint beamforming design and power allocation optimization has been formulated. To solve the formulated non-convex problem, we have applied AO technique based on DC programming and SCA to obtain a suboptimal solution. Numerical results show that the proposed algorithm can improve the performance of the MISO system based on transmissive RMS transmitter.
\ifCLASSOPTIONcaptionsoff
  \newpage
\fi

\bibliographystyle{IEEEtran}
\bibliography{reference}
\end{document}